\documentclass[apj]{emulateapj}

\usepackage{color}
\usepackage{graphicx,epstopdf}
\usepackage{enumerate}
\usepackage{natbib}
\usepackage[bookmarks=true, pdfstartview={FitH},hyperfootnotes=false]{hyperref}

\newcommand{\Chandra}{\emph{Chandra}}
\newcommand{\Suzaku}{\emph{Suzaku}}
\newcommand{\Rosat}{\emph{ROSAT}}
\newcommand{\XMM}{\emph{XMM-Newton}}
\newcommand{\IRAS}{\emph{IRAS}}
\def\arcdeg{\hbox{$^\circ$}}
\def\arcmin{\hbox{$^\prime$}}
\def\arcsec{\hbox{$^{\prime\prime}$}}

\shorttitle{Evidence for a very low-column density hole in the Galactic halo}
\shortauthors{Liu, Galeazzi, \& Ursino}

\begin{document}
\title{Evidence for a very low-column density hole in the Galactic halo in the direction of the high latitude molecular cloud MBM 16}


\author{W. Liu, M. Galeazzi, and E. Ursino}
\affil{Physics Department, University of Miami, Coral Gables, FL 33124, USA;
galeazzi@physics.miami.edu}



\begin{abstract}

Shadow observations are the only way to observe emission from the galactic halo (GH) and/or the
circumgalactic medium (CGM) free of any foreground contamination from local hot bubble (LHB) and solar
wind charge exchange (SWCX).
We analyzed data from a shadow observation in the direction of the high latitude, neutral hydrogen cloud
MBM 16 with \Suzaku. We found that all emission can be accounted for by foreground emission from LHB and
SWCX, plus power law emission associated with unresolved point sources. The GH/CGM in the direction of MBM 16
is negligible or inexistent in our observation, with upper limits on the emission measure of
$9.5\times 10^{-4}$ pc cm$^{-6}$ (90\% C.L.), at the lowest end of current estimates.

\end{abstract}



\section{Introduction}
\label{introduction}
Is the Milky Way surrounded by a hot halo a few kpc deep? Or is the X-ray emission we observe from a more
extended ($\sim100$ kpc) circumgalactic medium (CGM)? Is this region uniform of clumpy? Is its shape spherical
or pancake-like? These are some of the questions about the origin of the absorbed X-ray emitting region
at 3/4 keV observed by \Rosat. A region we know very little about, despite its influence on the
evolution of our Galaxy, due to its weak, diffuse emission and its overlap with foreground emission
from Solar Wind Charge eXchange (SWCX) and Local Hot Bubble (LHB).

The \Rosat\ All Sky Survey proved the existence of emission, characterized as
plasma at $\sim3\times10^{6}$ K. \Rosat\ showed that the emission is of
galactic origin (but farther away than the LHB) but could not give any better
indication of its location. More recently, higher resolution absorption and emission investigations
\citep[i.e.,][]{Nicastro02,Williams07,Galeazzi07,Gupta09b,Henley12}
detected the presence of O VII and O VIII lines at z$\sim$0, but the nature of this gas is still disputed.
One possible interpretation is that O VII and O VIII are associated to the predicted WHIM
\citep[e.g.,][]{Cen99} in the Local Group, thus providing a partial answer to the missing baryons problem.
On the other hand, this gas could be located well within the Galactic halo (GH). Distinguishing between
the two possibilities is not easy since detectors do not have high enough energy resolution. Following
\citet{Henley12}, we can define two models, the GH/CGM is a large shell of gas that envelopes the Milky Way
up to $\sim100$ kpc, with almost isotropic emission on large scale, or the GH/CGM is gas flowing away
from the Galactic Plane, at a distance of $\sim10$ kpc. In this case the emission should have radial
dependence $I=I_{0}cos|b|$, where $b$ is the galactic latitude.

Either model does not account for finer details in the GH/CGM structure. For example, if the cause of the
emission is due to hot gas infalling from our neighborhood, then we would expect a filament-like structure,
highly inhomogeneous. On the other hand, if the emission is due to hot gas escaping from the Milky Way,
then it is expected to be in the form of fountains, creating a patchy halo on smaller scale.

\citet{Henley12} analyzed a large number of O VII and O VIII observations performed with \XMM\ at different
positions in the sky. Their analysis ruled out the possibility that the gas is at $T\gg3\times10^{6}$ K,
but the high variance in the data does not allow to set stricter values on other parameters and to put
limits on the GH/CGM models: the temperature of the GH/CGM appears to be uniform
($\sim2\times10^{6}$ K) on all sky surveys, while the corresponding emission measure (EM) spans an order
of magnitude \citep{Henley13, Yoshino09}.
After filtering the emission line catalog by \citet{Henley10,Henley12}, \citet{Miller15} found a good
constraint using O VIII emission lines on the GH density profile, suggesting a galactic plane origin,
while the O VII constraint is dominated by temperature or density variations in the LHB.
However, the analysis was performed on single pointing observations,
giving rise to high uncertainties due to the poorly constrained contribution of the foreground (LHB and SWCX).

Shadow observations allow for the separation of foreground and background emission and are currently the only
tool to study the GH/CGM emission without the large systematic uncertainties coming from contamination
due to LHB and SWCX.

Shadow observations are performed by looking at a high column density cloud with distance of a few hundred
parsecs from the Sun, and at a low density target a few degrees away. Since the cloud is a strong absorber,
it shields most of the X-rays emitted by background sources and leaves an almost pure foreground spectrum.
The signal toward the low density line of sight, instead, includes also a large fraction of the contribution
from distant X-ray components. By comparing the spectra in the lines of sight of the cloud and the nearby
target, we can characterize the spectral properties of the distant component.

So far very few targets suitable for shadow experiments have been observed with any of the three major X-ray
satellites (\Chandra, \XMM, and \Suzaku). Our group has focused on the neutral hydrogen cloud MBM20 with
\XMM\ \citep{Galeazzi07} and \Suzaku\ \citep{Gupta09b}. In addition there are available observations of the
neutral hydrogen cloud MBM12 performed with \Chandra\ \citep{Smith05}, \Suzaku\ \citep{Smith07} and
\XMM\ \citep{Koutroumpa11}, and that of a relatively dense neutral hydrogen filament in the southern galactic
hemisphere \citep{Henley08}. A detailed comparison of the observations is reported in \citet{Gupta09b}.

Section \ref{DataReduction} of this paper will discuss the data reduction of
\Suzaku\ data. Section \ref{Results} will focus on the data analysis and results from the Shadow observation,
and section \ref{Discussion} will focus on comparison with other cloud observations (MBM12 and MBM20)
and the implications of the MBM16 result.

\section{Data Reduction}
\label{DataReduction}
\begin{deluxetable*}{lcccccc}
\tablewidth{0pc}
\tablecaption{Observations of MBM 16 On- and Off-cloud \label{table_obs}}
\tablehead{
\colhead{Target}           & \colhead{Observation ID}      &
\colhead{l (deg)}          & \colhead{b (deg)} & \colhead{Start Date} &
\colhead{Livetime (s)} & \colhead{N$_{H}$ (10$^{20}$ cm$^{-2}$}) }
\startdata
MBM 16 ON & 5080078010 & 170.58 & -37.28 & 2013-08-07 & 69053 & 33.2 \\
MBM 16 OFF& 5080073010 & 165.84 & -38.39 & 2013-08-09 & 70771 & 9.41 \\
\enddata
\end{deluxetable*}

The high-column-density molecular cloud MBM16 is part of a Suzaku SWCX key
project targets for the characterization
of SWCX, and has been observed several times in the last few years. During Suzaku cycle 8 we obtained the
observation of a low density region a few degrees away from MBM16, to be performed right after the key project
observation of MBM16. The timing of the two observations minimizes the SWCX variation and makes it possible to
model the SWCX as a constant component rather than being time variable. The most important parameter for the
observations and the analysis are reported in table \ref{table_obs}. In this paper we will
usually refer to the two targets as ``On-cloud'' and ``Off-cloud'' (or simply ``On'' and ``Off''),
meaning MBM16 and the low density target, respectively.

The assessment of the absorption properties is critical for this investigation.
Large scale neutral hydrogen surveys
\citep{Dickey90,Kalberla05}, however, have very poor angular resolution, of the
order of the \Suzaku\ field of view, and cannot resolve the individual parts of the two targets. On the other hand,
\IRAS\ 100 $\mu$m maps have excellent angular resolution and allow for a full characterization of the targets.
In order to compute the required hydrogen density, we converted the \IRAS\ 100
$\mu$m intensity to neutral hydrogen column  density (N$_{H}$) using
the ``typical'' high-latitude 100$\mu$m/N$_{H}$ ratio of $0.85\times10^{-20}$
cm$^{2}$ MJy sr$^{-1}$ \citep{Boulanger88}.

The data reduction procedure is relatively standard and we already used it several times for other works
\citep[e.g.,][]{Mitsuishi12}, including the analysis of the Suzaku key project data. The raw data are reprocessed
with the most update calibration files. The 5x5 and 3x3 mode-data are merged to increase the size of the sample.
We filtered the data for high flaring and anomalous regions where the satellite is exposed to high particle
flux and extracted clean event files and the 0.4-2.0 keV maps of the targets. We identified and removed
point sources using wavelet detection for a 120$\arcsec$ PSF. After additional filtering for the cut-off rigidity
of the Earth's magnetic field, we extracted the spectra and generated the corresponding exposure maps,
response and ancillary files, and spectra of non-X-ray background. The X-ray and non-X-ray spectra are normalized
to a uniform 20$\arcmin$ circular field. The non-X-ray spectrum is further rescaled to match the observed spectrum
in the 11-14.5 keV range, where the effective area is negligible and the observed signal is only
of non-X-ray origin. We did not include data from the XIS0 and XIS3 chips. They have a small effective area
at the low energy tail of our interest and give little contribution to the signal.

In recent years a new contaminant is increasingly affecting observations, the O I line at 0.525 keV
\citep{Sekiya14} resulting from fluorescence of solar X-rays with neutral O in
the Earth's atmosphere. Lack of proper handling of the O I signal leads to large bias in the estimate of the
nearby O VII lines and could lead to wrong interpretation of the properties of the emitting plasma. Rather than
excluding time intervals with elevation angle larger than 60$\arcdeg$ as
suggested by \citet{Sekiya14} and losing a lot of the available data, we included the O I contamination in
our models by adding an emission line at 0.525 keV.

\section{Data Analysis and Results}
\label{Results}

The X-ray emission toward MBM16 On- and Off-cloud should consist of foreground emission (LHB plus SWCX),
GH/CGM emission, and extragalactic X-ray background from unresolved point sources. To model the foreground
emission (LHB and SWCX), a single apec thermal model has often been used in the past
\citep{Foster12,Galeazzi07,Henley08,Gupta09b}. In our analysis,  we used an improved model that includes
an un-absorbed apec model with O abundance fixed at 0, plus three $\delta$-functions at the energies of O VII
line (0.57 keV), and O VIII line (0.65 keV and 0.81 keV) to take into account the variation induced by SWCX.
We also included one additional Gaussian line at energy of 0.525 keV to model the O I line as mentioned in Section
\ref{DataReduction}. We modeled the GH emission as an absorbed apec thermal model, with its temperature and
normalization free, and modeled the extragalactic emission with an absorbed power law, where both the index and
normalization are free parameters.

In order to separate the foreground and background emission, we fitted the On- and Off-cloud spectra
simultaneously, with the model parameters of the On- and Off-cloud tied together except for the
absorbing column density. We performed
spectral analysis in the energy band 0.4-5.0 keV, using the XSPEC version 12.8
\citep{xspec}, adopting the metal abundance model by \citet{Anders89}. Both spectra are re-binned to have at
least 50 counts per bin. To verify that the fit had not become trapped in a local minimum, we explored the
local parameter space by varying individual parameters over a range centered on the best-fitting values.

\begin{figure}
\plotone{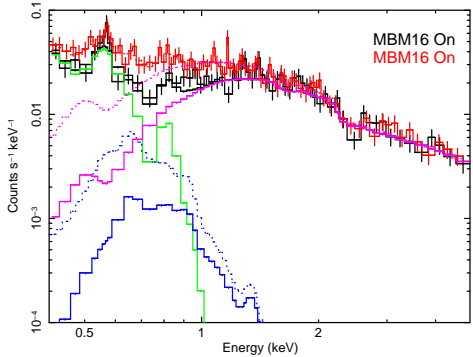}
\caption{The spectra of MBM 16 On- (black) and Off-cloud (red), along with the best
fitting model. We show the decomposition of the best fitting model into its various components:
the foreground emission (in green), the GH/CGM emission (in blue), and the extragalactic emission from unresolved
point sources (in magenta). The solid lines are for the On-cloud, and the dotted lines are for Off-cloud.
\label{spectrum}}
\end{figure}

In Figure \ref{spectrum} we show the On- and Off-cloud spectra of MBM16 together with the best
fitting model, which is decomposed into different model components (foreground in green, GH in blue
and extragalactic component in magenta). In general the fit is good with a reduced $\chi^{2}$ of 0.93
($\chi^{2}/dof=302.61/326$). In table \ref{table_result} we present the results of spectral modeling
(errors are quoted at 90\% confidence). The best fit temperature of the LHB is consistent with
previous studies \citep{Snowden98, Kuntz00, Liu15b}. We also verified that the contamination from O I
line to the O VII line is small
and excluding the O I component does not affect our results. The temperature of the GH/CGM is consistent
with previous estimates \citep{Henley13}, but with large uncertainty. This is because
the GH/CGM component is negligible comparing to other components (foreground and extragalactic background)
as seen in Figure \ref{spectrum}.

\begin{figure}
\plotone{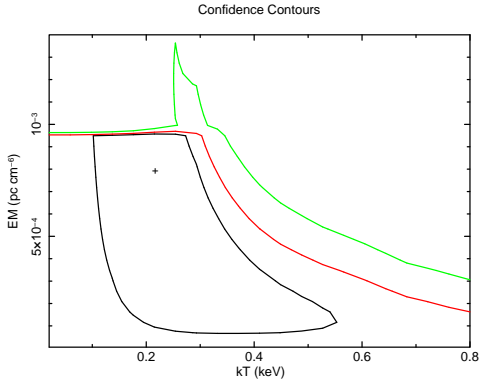}
\caption{The GH temperature and emission measure constraints at the 68\% (black), 90\% (red), and 99\% (green)
confidence levels from the simultaneously fitting of MBM 16 On- and Off-cloud.
\label{contour}}
\end{figure}

\begin{figure*}
\plotone{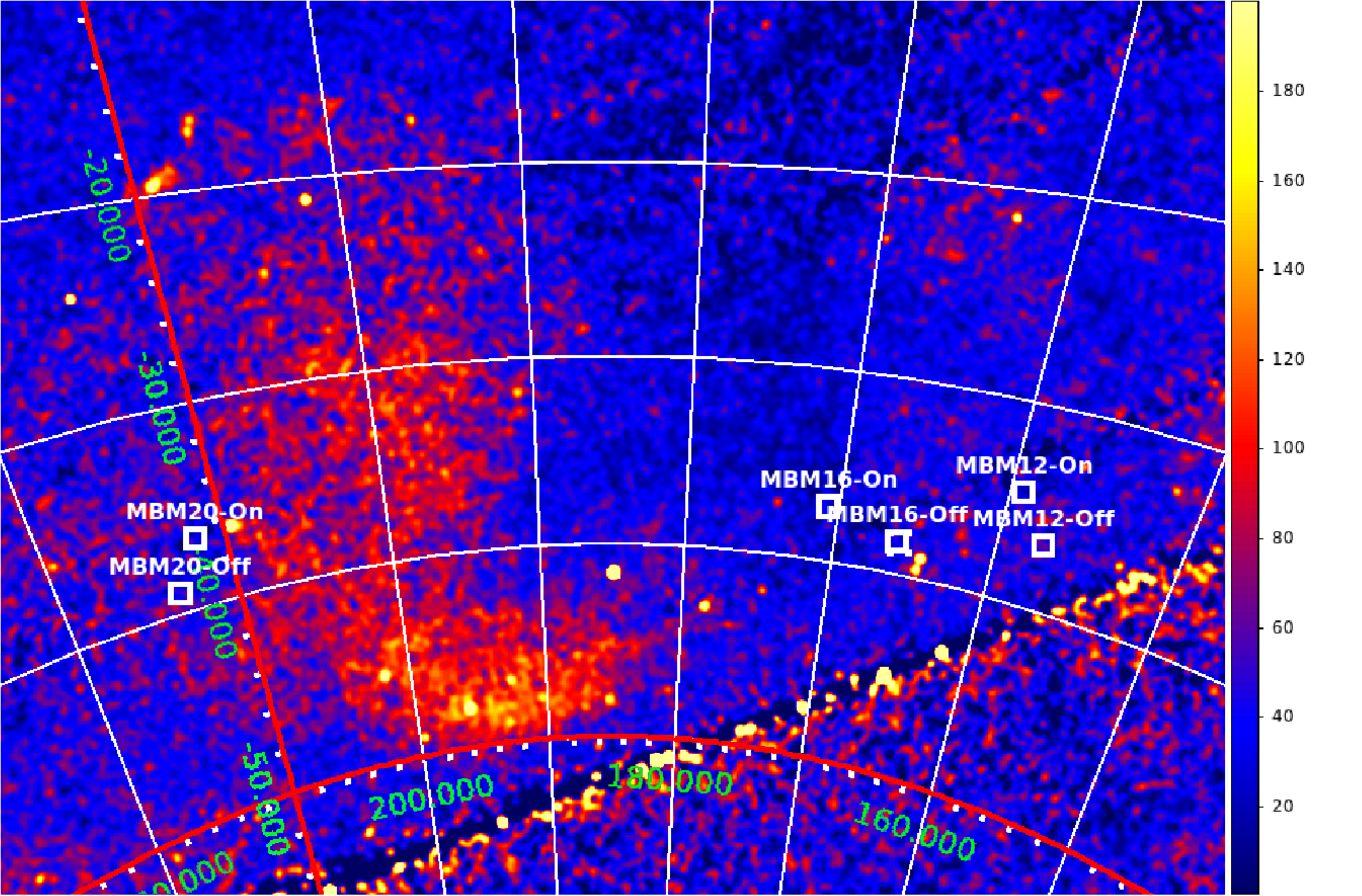}
\caption{The squares ($1\arcdeg\times1\arcdeg$, for plotting purpose) show locations of three shadow targets,
MBM 12, MBM 16, and MBM 20, in \Rosat\ All-sky Survey map in R4 band (units $10^{-6}$ counts s$^{-1}$
arcmin$^{-2}$, the grid is in Galactic coordinate).
\label{rosat_targets}}
\end{figure*}

\begin{deluxetable*}{lccccccccccc}
\tablecolumns{12}
\tablewidth{0pt}
\tabletypesize{\scriptsize}
\tablecaption{Spectral Fitting Result \label{table_result}}
\tablehead{
    \multicolumn{2}{c}{LHB} & \multicolumn{3}{c}{SWCX}  &
    \colhead{} & \multicolumn{2}{c}{GH} &
    \multicolumn{2}{c}{Power Law} & & \multicolumn{1}{c}{$\chi^{2}$/dof} \\
\cline{1-2} \cline{3-5} \cline{7-8} \cline{9-10} \\
\colhead{T}   & \colhead{EM}  & \colhead{O VII}       &
\colhead{O VIII\tablenotemark{a}}      & \colhead{O VIII\tablenotemark{b}}      & \colhead{} & \colhead{T}    &
\colhead{EM}   & \colhead{$\Gamma$}    & \colhead{Norm\tablenotemark{c}}           &
\colhead{O I}         &  \\
\colhead{keV}         & \colhead{pc cm$^{-6}$}   &  \colhead{(LU)}  &
\colhead{(LU)}        & \colhead{(LU)}     & \colhead{}                &  \colhead{keV}   &
\colhead{$10^{-3}$ pc cm$^{-6}$}  &                   &                  &
\colhead{(LU)}                    & }
\startdata
$0.09^{-0.04}_{+0.03}$ & $0.03^{-0.01}_{+0.12}$  & $3.78^{-0.87}_{+0.80}$ &
$0.86^{-0.39}_{+0.40}$ & $0.34^{-0.20}_{+0.19}$  &  & $0.22^{-0.20}_{+0.44}$ &
$0.79^{-0.79}_{+0.18}$ & $1.58^{-0.08}_{+0.08}$   & $1.03^{-0.06}_{+0.06}$ &
$0.83^{-0.83}_{+0.65}$ & $302.61/326$ \\
\enddata
\tablecomments{Errors are quoted at 90\%\ confidence.}
\tablenotetext{a}{O VIII line at 0.57 keV.}
\tablenotetext{b}{O VIII line at 0.81 keV.}
\tablenotetext{c}{Normalization of power law fit at 1 keV in units of
$10^{-3} $\textrm{photons keV$^{-1}$ s$^{-1}$ cm$^{-2}$}. }
\end{deluxetable*}

In fact, the magnitude of the GH/CGM component is consistent with zero. Figure \ref{contour} shows the statistical
relevance of the GH/CGM component as function of temperature and emission measure. The contour plot shows the
68\%, 90\%, and 99\% confidence region as black, red, and green lines respectively.
At 90\% confidence levels, the upper limit of emission
measure is $\sim9\times 10^{-4}$ pc cm$^{-6}$, while the lower limit is well consistent with zero. We tested the case
with zero contribution from the GH/CGM component in our model. From the f-test, there is a zero probability
significance of improving the fit by including a GH/CGM contribution. Our results are therefore consistent
with no GH/CGM emission in the direction of MBM16.

We also verified the effect of the foreground model on the statistical significance of our model.
For example, we used the AtomDB Change Exchange (ACX, \citet{Smith12}) model to represent the SWCX component,
instead of the Gaussian lines. We found that the model used to represent the foreground does not affect
our result, and all our tests are consistent with a negligible contribution from GH/CGM.

\section{Discussion}
\label{Discussion}
Multiple authors have investigated the X-ray emission of empty fields in the sky to extrapolate the
contribution from GH/CGM. Those papers show significant variation between pointings, including fields
where no significant emission was detected, perhaps an indication of a patchy medium. However, those
results are significantly affected by uncertainty in the removal of background emission, making any strong
conclusion difficult.

The current result shows, without significant systematic uncertainty, that the GH/CGM emission in the direction of MBM 16 is
quite small or inexistent, with an upper limit on the plasma emission measure at the lowest end of the
sample covered by \citet[Figure 6]{Henley13}. Using simple assumptions, this result can be
converted to significant upper limits on the density of the medium, depending on the model
used. Specifically, assuming a uniform halo, the 90\% C.L. upper limit of
$9.5\times10^{-4} $ pc cm$^{-6}$ for the emission measure would translate to a maximum density
of $9.7\times10^{-5}(Z/Z_{solar})^{-1}$ cm$^{-3}$ for an extended (100 kpc) halo, and
$4.4\times10^{-4}(Z/Z_{solar})^{-1}$ cm$^{-3}$ for a more compact (5 kpc) one. While the statistical uncertainty on this
result is still not sufficient to discriminate between GH/CGM models, it places the upper limit on emission
measure at the lower end of prediction from uniform halo models, confirming the possibility of a
non-uniform halo.

The number of shadow observations currently available is very limited, making any
general conclusion limited at this time. A summary of such investigations is available in \citep{Gupta09b}.
However, three observed shadow targets
are closely spaced together (MBM16, MBM12, and MBM20 see Figure \ref{rosat_targets}) and can be compared
directly. Of the three, only MBM16 does not show any GH/CGM emission, which would indicate a highly
inhomogeneous medium.
However, a closer look at the measurements in the direction of MBM12 and MBM20 may show a different picture.
MBM12 (observed by \citet{Smith07})
is very close to MBM16, and any significant difference between the halo emission in the two directions
would be strong evidence of a highly inhomogenous medium.
However, \citet{Koutroumpa11} has shown that the MBM12 observation was affected by strong
variation in SWCX between on and off-cloud, making any conclusion impossible. It is conceivable, indeed,
that the measured GH/CGM emission measured in the direction of MBM12 was, in fact, the result of SWCX, and
that the GH/CGM emission was compatible with MBM16.

MBM20 is slightly further away, and its positive emission would point to a variation in GH/CGM on the scale
of some degrees. However, MBM20 is at the edge of the Eridanus enhancement
\citep{Burrows93,Snowden95}, which again could mimic
any emission from GH/CGM. In conclusion, the current, very limited evidence, is
in fact consistent with a region of very low, or non-existent emission from GH/CGM.

\acknowledgments

This work was funded by NASA grant NNX11AF80G and NNX13AI04G.

\end{document}